\documentstyle[epsfig]{aipproc}

\begin{document}

\title{X-ray Selected BL Lacs and Blazars}

\author{Eric S. Perlman}

\address{Space Telescope Science Institute}

\maketitle

\begin{abstract}

With their rapid, violent variability and broad featureless continuum
emission, blazars have puzzled astronomers for over two decades.
Today blazars represent the only extragalactic objects detected in
high-energy gamma-rays.  Their spectral energy distributions (SEDs)
are characteristically double-humped, with lower-energy emission
originating as synchrotron radiation in a relativistically beamed jet,
and higher-energy emission due to inverse-Compton processes.  This has
accentuated the biases inherent in any survey to favor objects which
are bright in the survey band, and should serve as a cautionary note
both to those designing new surveys as well as theorists attempting to
model blazar properties.  The location of the synchrotron peak
determines which blazar population is dominant at GeV and TeV
energies.  At GeV energies, low-energy peaked, high luminosity
objects, which have high $L_C/L_S$ ratios, dominate, while at TeV
energies, high-energy peaked objects are all that is seen.  I review
the differences between low-energy peaked and high-energy peaked
blazars, and models to explain those differences.  I also look at
efforts to bridge the gap between these classes with new surveys.  Two
new surveys have detected a large population of high-energy peaked
emission line blazars (FSRQ), with properties somewhat different from
previously known objects. This discovery has the potential to
revolutionize blazar physics in a way comparable to the discovery of
X-ray selected BL Lacs ten years ago by {\it Einstein}.  I cull from
the new and existing surveys a list of $z<0.1$ high-energy peaked
blazars which should be targets for new TeV telescopes.  Among these
are several high-energy peaked FSRQ.

\end{abstract}

\section{Introduction}

Blazars have the most extreme properties of any class of active
galactic nuclei.  In every wavelength range, their properties
are dominated by a broad, highly variable continuum.
This continuum has a characteristic, double-humped shape (Figure 1),
indicative of two emission processes.  At lower energies, synchrotron 
radiation dominates the energy budget, but at X-ray through
gamma-ray energies, inverse-Compton processes increasingly dominate
the properties we observe.  The rapid, violent variability that is the
hallmark of these objects (blazars can vary in brightness by factors
of ten or more, and doubling on timescales of hours is seen in their
lightcurves; see the review herein by Rita Sambruna \cite{Sambruna99}),
forces us to explain their properties as a consequence of viewing a
relativistic jet moving very close to our line of sight (see
\cite{UrPa95,Koll94} for reviews).

\begin{figure}

\centerline{\epsfig{file=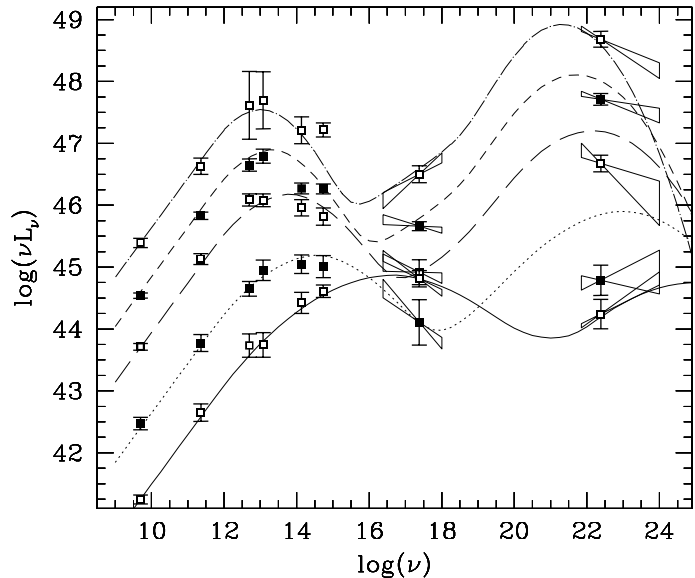,height=3in}}

\caption{The variation of the average spectral energy distribution 
of blazars with radio luminosity [9].  The low-energy component, due
to synchrotron radiation, peaks in the infrared for ``red'',
low-energy peaked blazars, and at UV/X-ray energies for ``blue'',
high-energy peaked blazars.  Note how the location of the synchrotron
peak varies with luminosity.}

\end{figure}

The blazar class covers a very wide range in luminosity as well as
peak frequency.  More luminous objects tend to peak at lower frequencies,
but there is a wide scatter in this relation \cite{Fossati98}.
Historically, optical spectroscopic properties have been used to
separate blazars into two divisions: flat-spectrum radio quasars
(FSRQ) have strong, broad emission lines, while BL Lacs have very
faint or no emission lines.  However, this distinction now appears  
arbitrary, as recent work has
shown a continuous distribution of emission line luminosities and
equivalent widths \cite{ScFa97}.

I will review the surveys which have been used to find blazars as well
as their biases, and show how this has produced two populations
with somewhat different properties.  I will for pedagogical
reasons adopt the traditional division between BL Lacs and FSRQ, but I
believe that one of the most important tasks the new surveys must
undertake is to define new, physically based classes for blazars.  I
will describe the latest crop of surveys and their findings, and
``round up'' a herd of objects which should be targets for the
new generation of VHE gamma-ray observatories.

\section{Survey Methods and Biases}

Because of their rareness, blazars have an unfortunate history
of divisions ``invented'' because of observed
properties or selection methods which may not have any physical
basis. The result has been confusion not only over how to define
subclasses and their properties, but indeed over the
definition of the blazar class itself!  The BL Lac/FSRQ division
is an example of this phenomenon; another  
is the definition of ``radio-selected'' and ``X-ray selected''
BL Lac classes, based on the survey in which an object was found.
Yet several well known objects turn up in both
radio and X-ray surveys, for example Mkn 421, Mkn 501 and BL Lac.

Our understanding is helped considerably if we take a step back and
try to understand the biases inherent in single-band surveys.
The key point (which seems obvious but is in fact surprisingly subtle)
is that any survey selects preferentially objects that are bright in
the survey band.  Thus the overwhelming majority of blazars selected
in X-rays peak in the UV or X-rays, while nearly all blazars
selected in the radio peak at much lower (IR) energies (Figure 1).
The subtlety lies in the fact that these two methods attack
opposite ends of parameter space (Figure 2), and {\it do} seem to find
objects with somewhat different properties (more about this in
\S III).  Thus while inquiries into blazar properties have achieved
much by using X-ray and radio selected samples, they have hardly
delved into what connects them.

\begin{figure}
\centerline{\epsfig{figure=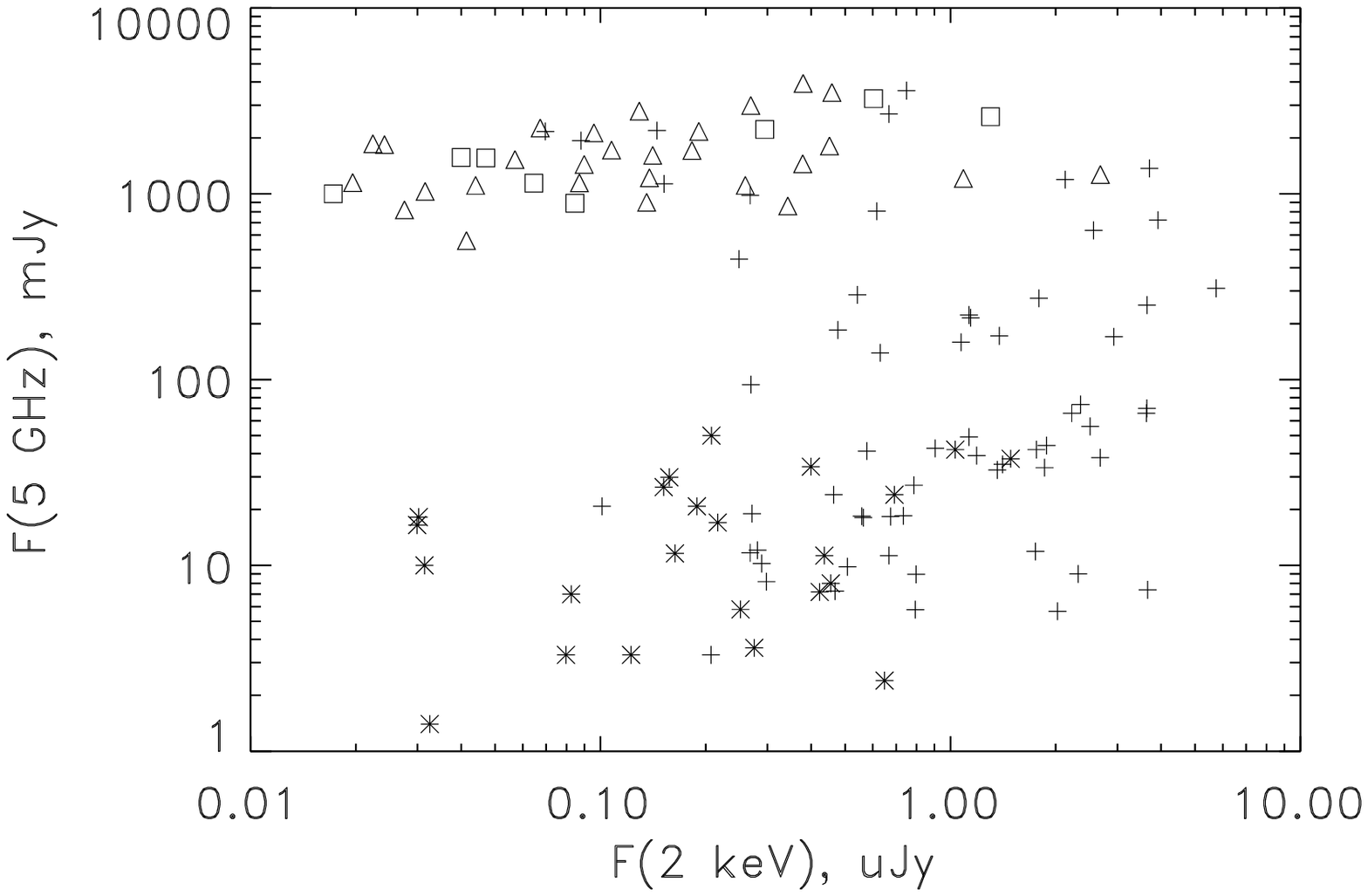,width=3in}
\epsfig{figure=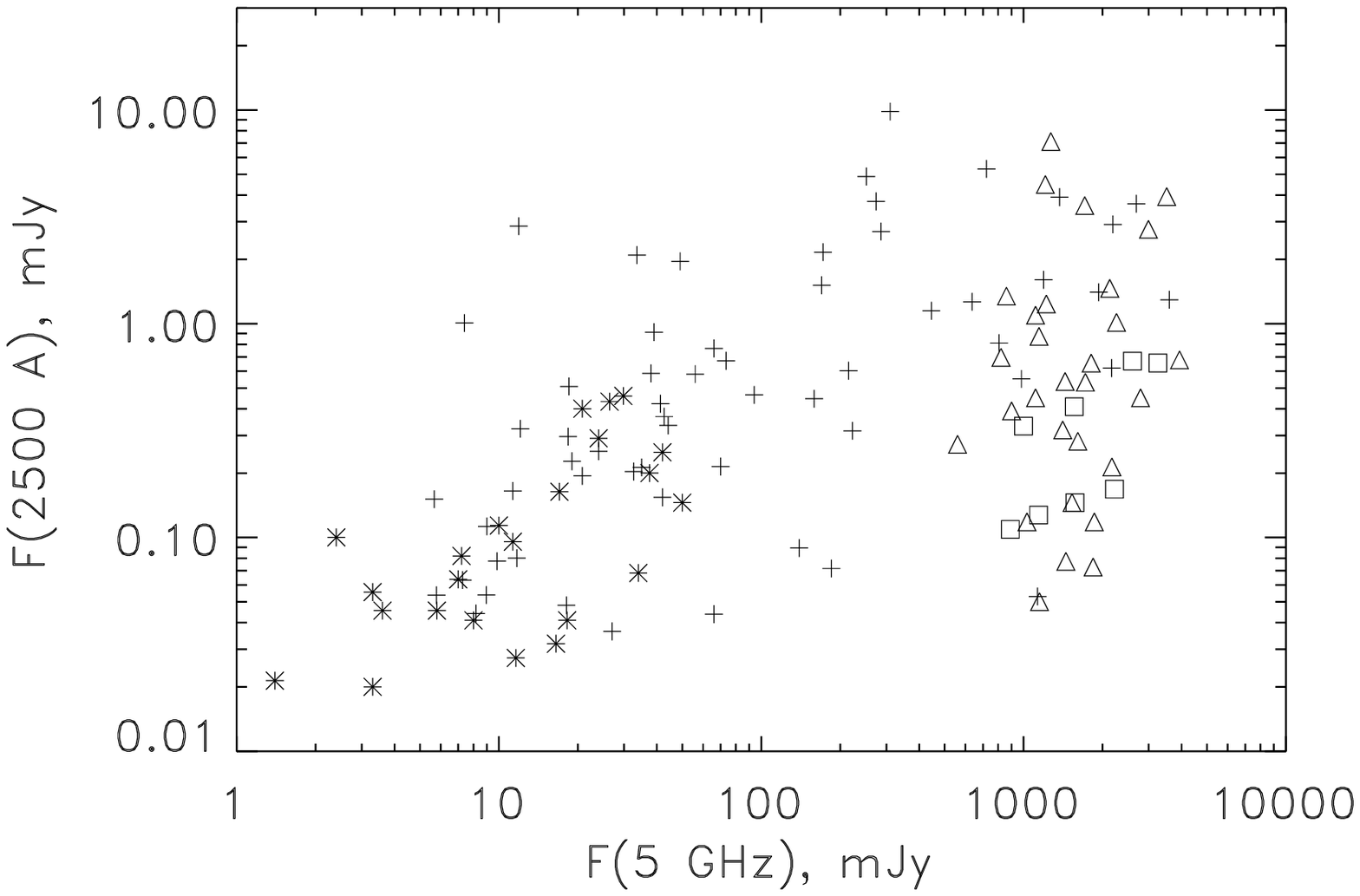,width=3in}}
\caption{Two views of the radio-optical-X-ray flux parameter space for
previous samples of BL Lacs.   1 Jy BL Lacs are shown 
plotted as triangles, S4 BL Lacs are shown plotted as squares, Slew
BL Lacs are shown as pluses, and EMSS BL Lacs are shown as asterisks.
Note the large gaps which existed prior to 1995 in our coverage of 
this parameter space.}

\end{figure}

Today we speak of ``high-energy peaked'' and ``low-energy peaked''
blazars, referring to objects which peak at (respectively) UV/X-ray or
infrared energies.  The disparity in the peak frequencies indicates
significant differences in jet physics.  This is expected on
theoretical grounds, since the characteristic electron energy for
synchrotron emission is directly related to the magnetic field
$(\gamma_{peak} \propto \sqrt{\nu_{peak}/B})$.  Moreover, the trends
we find with luminosity (decreasing $\nu_{peak}$, increasing emission
line luminosity) indicate substantially more cooling in more
luminous objects.

The location of the synchrotron peak is intimately connected to which
kinds of objects are observed to dominate at GeV and TeV gamma-ray
energies.  At GeV energies, lower-energy peaked, high-luminosity
objects dominate, as they have much higher ratios $L_C/L_S$.  These
objects, however, do not make electrons with $\gamma \sim 10^{6-7}$,
which are required for X-ray synchrotron emission, probably because of
increased cooling.  Thus at TeV energies, objects which peak in the
UV/X-rays (i.e., high-energy peaked or X-ray selected blazars) are all
that is seen.

Unfortunately, current surveys do not contain enough information to
tell us which kind of blazar (high or low-energy peaked) is more
common.  This is because current complete samples cover very shallow
dynamic ranges (Figure 2).  There is a general indication that
high-luminosity objects are less common \cite{UrPa95}.  However,
deeper surveys are needed, because finding the absolute number of
either kind of object requires correcting current surveys for the
objects it does {\it not} find -- and to do this, we must go deep
enough in both radio and X-ray so that radio surveys start detecting
significant numbers of high-energy peaked blazars, and vice versa.

\section{The Properties of Red and Blue Blazars}

Over the last decade, many workers have delved into the properties of
high-energy peaked and low-energy peaked blazars, by using samples of
BL Lacs selected in the radio and X-rays.  BL Lacs were used in this
work because until very recently there were no FSRQ known to peak at
UV/X-ray energies(\cite{Perlman98,PaGi97,PaPe99}; \S IV).  These works
found significant differences between the properties of blue,
high-energy peaked BL Lacs (HBLs) and red, low-energy peaked BL Lacs
(LBLs):

\begin{itemize}

\item HBLs are less luminous in radio and
bolometrically \cite{Sambruna96,Fossati97,Fossati98}.

\item HBLs are less core-dominated in the radio than LBLs 
\cite{PeSto93,PeSto94,L-M93,Koll96,Rector99}.

\item HBLs are less polarized than LBLs, with a smaller duty cycle, 
and tend to have a preferred position angle of polarization, while 
LBLs do not \cite{Jannuzi93}.

\item Occupy a different region of X-ray-optical-radio parameter space (Figure
2), and in fact a unique region of parameter space in X-ray-optical
and radio-optical spectral index space (Figure 3, \cite{Stocke91}).

\item HBLs tend to have steeper X-ray and optical-X-ray continua than LBLs
\cite{Perlman96,Sambruna96,Urry96,Lamer97,Padovani97}.

\item HBLs are distributed differently in space, with more objects or 
more luminous objects (current samples cannot discriminate 
between these possibilities) at low redshifts  
\cite{Morris91,Wolter94,Perlman96,Bade98,Giommi99,Rector99}; 
while LBLs are consistent with either
a uniform distribution with redshift or more objects at high redshift
\cite{Stickel91}.

\end{itemize}

As with their properties, the relationship between HBLs and LBLs has
been a subject of active debate in the literature.  At first it was
thought that they were related through viewing angle (e.g.,
\cite{Ghis93} and refs. therein).  This explained many properties in a
natural way, for example the differences in polarization behavior and
radio core dominance (though see \cite{Rector99} for new
counter-evidence), as well as the observed difference in space density
(which could have been a selection effect, however; see \S II).  A
second model was proposed by Padovani \& Giommi \cite{GiPa94,PaGi95},
under which HBLs and LBLs represent two ends of a continuous
distribution of synchrotron peak frequencies.  It turns out that both
descriptions have problems.  Rita Sambruna showed in her thesis
\cite{Sambruna96} that differences in in viewing angle cannot produce
a variation of $10^4$ in peak frequency.  And while the ``different
peak frequencies'' description is accurate phenomenologically, it
cannot by itself explain the differences in radio core-dominance
\cite{Koll96,Stocke96} and polarization PA \cite{Stocke96} behavior.

This question is still open, but a modern view is evolving which
basically says that both the viewing angle and different spectral
energy distributions pictures have a piece of the puzzle.  Two competing
models now ascribe the HBL-LBL relationship to combinations of
luminosity and viewing angle \cite{GeoMa98}, or luminosity and peak
frequency \cite{Ghis98}.  Current data cannot distinguish
between these models, although further investigation of the
polarization differences with larger samples and in multi-wavelength
campaigns, offer in my view the best hope for doing so.

\section{The New Surveys: Bridging the Gaps}

As I discussed in \S II, the existing complete samples of blazars
suffer from several problems.  First of all they are small: typically
a few dozen objects at most.  There are also various concerns about
completeness, particularly at the lowest luminosities (e.g.,
\cite{BroMar93,MarBro95,MarBro96,Perlman96,RSP99}).  But the most 
difficult problems have to do with the small dynamic ranges 
covered in flux and luminosity (cf. Figure 2).  

In this section we review the latest information on existing samples.
However, we will concentrate largely on four new surveys which are
bridging the gaps in our coverage of parameter space.  These surveys
are allowing us to for the first time actively pursue the connections
between blazar classes and get at the real physics.  The most exciting
discovery of these surveys is the existence of a large population of
high-energy peaked FSRQ.  We will reanalyze two existing samples in
the light of these findings, and show that their makeup is consistent
with the new surveys.

The existing samples of blazars are listed below:

\begin{itemize}

\item {\it Einstein} Slew Survey \cite{Perlman96b,Perlman99}:  $F(X) \gtrsim 10^{-11}
{\rm ~erg ~cm^{-2} ~s^{-1}}$, $F(R) > 1$ mJy, 50\% of the sky.  66 BL
Lacs, 19 FSRQ.  Includes all known TeV emitters.

\item {\it Einstein} EMSS \cite{Stocke91,Morris91,Rector99,Perlman99}: $F(X) > 2 \times 
10^{-13} {\rm ~erg ~cm^{-2} ~s^{-1}}$, $F(R) > 1$ mJy, 2\% of the sky.
43 BL Lacs, 16 FSRQ.

\item 1 Jy \cite{Stickel91,Stickel93,Stickel94}: $F(R) > 1$ Jy, 60\% of sky.
37 BL Lacs, 222 FSRQ, but not completely identified.

\item S5 \cite{StiKu96}:  $F(R) > 250$ mJy, $\delta > 70^\circ$.  11 BL Lacs,
20 FSRQ, but not completely identified.

\item S4 \cite{StiKu94}: $F(R) > 500$ mJy, $35^\circ < \delta < 70^\circ$.
7 BL Lac, 56 FSRQ, but not completely identified.

\end{itemize}

\begin{figure}

\centerline{\epsfig{file=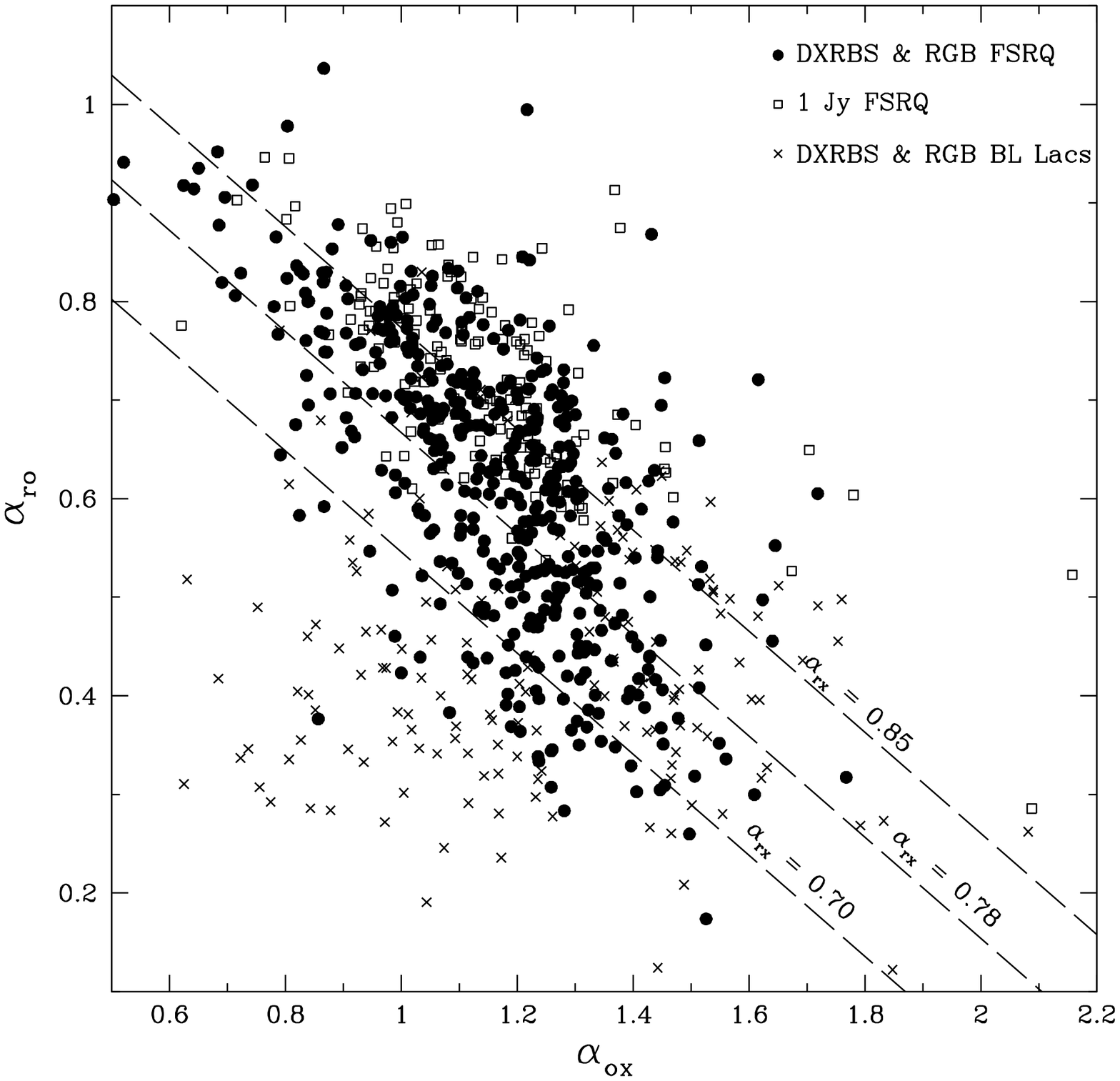,width=3.5in}}

\caption{X-ray-optical and Optical-Radio Spectral indices of
FSRQ and BL Lacs discovered in radio and X-ray surveys[34,35].  Note the
vastly different ranges of parameter spaces covered by historical
radio surveys (1 Jy: open squares) compared to X-ray surveys (DXRBS
and RGB: see \S 4).  Prior to 1996, in fact, the parameter space
covered by X-ray and radio techniques was almost {\it completely}
disjoint, as shown in Figure 2.}

\centerline {\epsfig{file=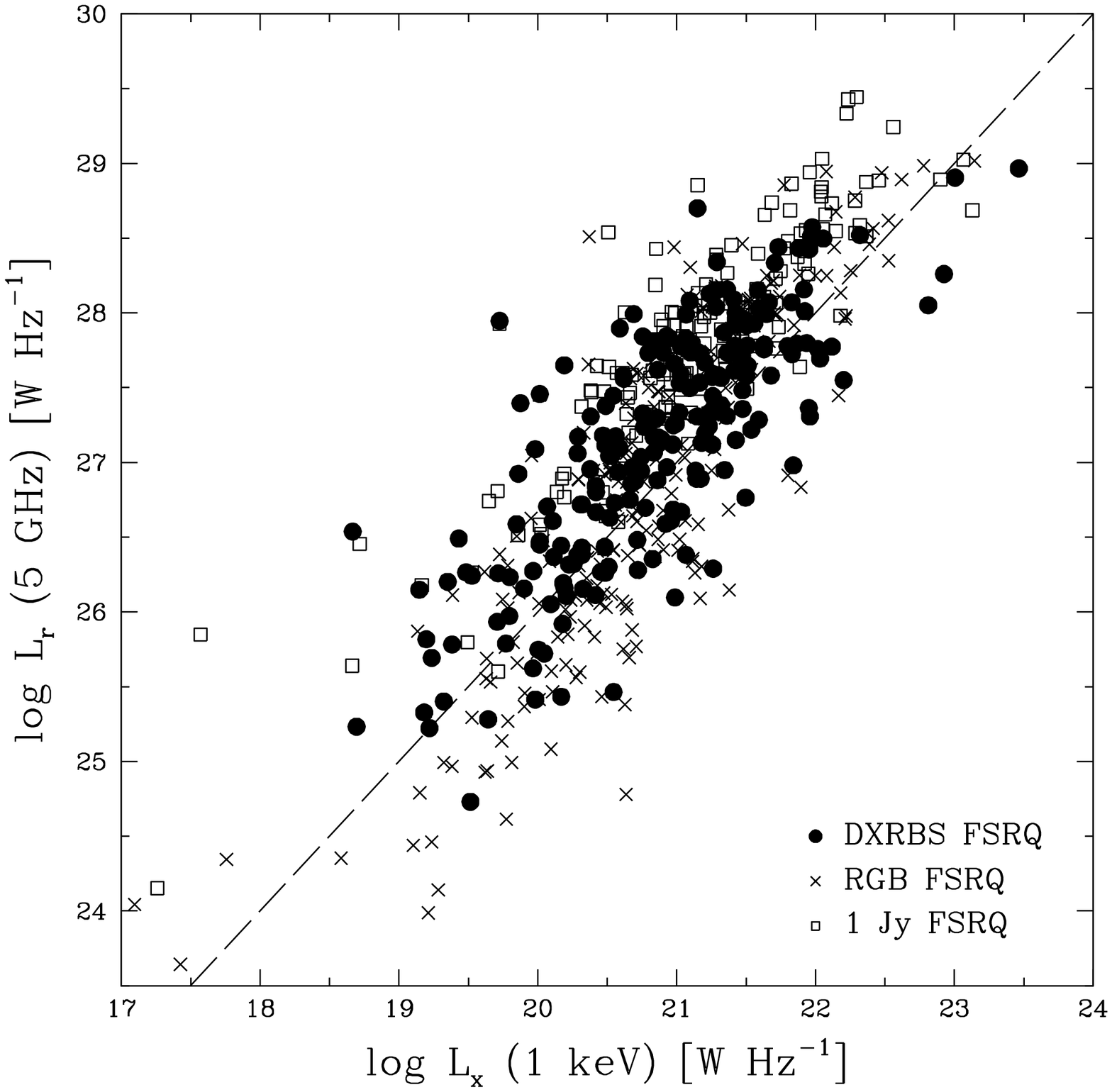,width=3.5in}} 

\caption{X-ray and Radio luminosities for FSRQs[34,35].  Note
that high-energy peaked FSRQs (right of the dashed line) are
less luminous in the radio (for the same X-ray luminosity).  DXRBS and
RGB sample unexplored regions of parameter space,
containing objects $\sim 20 \times$ less luminous than previous surveys.}

\end{figure}

Four new surveys are in progress, filling vast new
regions of parameter space (Figures 3,4).  These surveys are
listed in Table 1.

\begin{table}[ht!]
\caption{New Surveys for Blazars} 
\begin{tabular*}{6.35in}{lccccccl}
& & & & &  \multicolumn{2}{c}{Flux Limit} \\
Survey 	      &  BL & $z<0.1$ &FSRQ &  $z<0.1$ & F(0.1-2 keV) & F(R) & Types of Objects \\ 
& Lacs &&&& ${\rm ~erg ~cm^{-2} ~s^{-1}}$ & mJy & \\
\tableline
DXRBS\cite{Perlman98,Landt99,PaPe99} &  40 & 5\tablenote{36 BL Lacs with redshifts.}& 218\tablenote{IDs nearly complete; 49 HBL-like FSRQs with $\alpha_{rx}< 0.78$ and $\alpha_{ro} < 0.6$.} &  0 &  $10^{-14}$ & 25-50 & most LBL/intermediate \\
RGB \cite{Brink97,L-M98,L-M99} & 127 & 14\tablenote{49 BL Lacs with redshifts.} & 252\tablenote{IDs nearly complete; 96 HBL-like FSRQs with $\alpha_{rx}< 0.78$ and $\alpha_{ro} < 0.6$.} & 9 & $10^{-12}$ & 25 &  most HBL/intermediate \\
NVSS-RASS\cite{Giommi99} & 58\tablenote{Spectroscopy ongoing: 155 candidates, 85\% efficiency expected.} &  1\tablenote{36 with redshifts}	& ? & ?	& $10^{-12}$  & 3.5 & all HBL \\
REX\cite{Caccia99} & ?\tablenote{Candidate list not released; surveys ROSAT pointed database} & ? & ? & ? & $10^{-14}$ & 3 & mostly HBL \\
\end{tabular*}
\end{table}

The depth and size of these new surveys has allowed them to probe much
deeper into the luminosity function of blazars than ever before.
Close examination of Figures 3 and 4, and comparison with Figure 2
reveals two key discoveries.  First of all, the number of FSRQ with
radio luminosity $L_R < 10^{26.5} {\rm ~W ~Hz^{-1}}$ has increased
nearly ten-fold, and for the first time luminosities below
$10^{25.5}{\rm ~W ~Hz^{-1}}$ are being reached (an equally large
expansion of the number of low luminosity BL Lacs is also taking
place).  This is important because the knee in the radio luminosity
function of FSRQ is located at or near $10^{26.5} {\rm ~W ~Hz^{-1}}$,
and the location of the knee and the shape of the luminosity function
below the knee are very poorly constrained due to the paucity of low
luminosity objects in current samples.  The second, equally important
discovery, is that because these surveys have plugged the holes in our
coverage of X-ray-optical-radio parameter space, they have revealed
large numbers of blazars in regions of parameter space where
previously very few were known \cite{Perlman98,L-M98,PaPe99,L-M99}.
These objects fall into two categories: intermediate BL Lacs, which
have peak energies $\sim 1-10$ eV, and X-ray bright FSRQ, which some
authors had predicted did not exist due to the observed continuity in
broadband spectral properties between the blazars known in complete
samples in 1996 \cite{Sambruna96}.  These two discoveries are in fact
not completely independent of one another, because the lowest
luminosities are dominated by high-energy peaked objects (Figure 4).

The X-ray bright FSRQ are in fact particularly important for our
knowledge of the class, because they overlap significantly in radio
and bolometric luminosities with the low-energy peaked BL Lacs.
Investigations into their properties and comparisons with previously
known FSRQ hold the promise of truly understanding the connections
between different classes of blazars.  A similar thing can be said for
the intermediate BL Lacs; but since the range of parameter space
opened up is larger for the emission line objects and the potential
impact on VHE gamma-ray astronomy from them is greater, I will
concentrate on them in this paper.

The DXRBS collaboration has begun an investigation into the properties
of X-ray bright FSRQ.  We find that these objects differ in important
ways from their lower-energy peaked cousins (in most cases paralleling
radio/X-ray selected BL Lac differences).  These differences (Figure
5) include ROSAT spectra that are steeper than known FSRQ
\cite{PaPe99,Sambruna97}, and SEDs that appear to peak in the
UV/X-rays, based on ROSAT, optical and radio catalog data
\cite{PaPe99}.  They also overlap significantly with HBL BL Lacs on
the $(\alpha_{ro},\alpha_{ox})$ plane (Figure 3).  However,
observations of several of these objects at harder X-ray energies with
ASCA and SAX have revealed that most (but not all) have rather flat
spectra, more similar to other FSRQ than HBL BL Lacs
(\cite{Sambruna99,Costa99}, although note that the selection criteria
were different).

\begin{figure}
\centerline{\epsfig{file=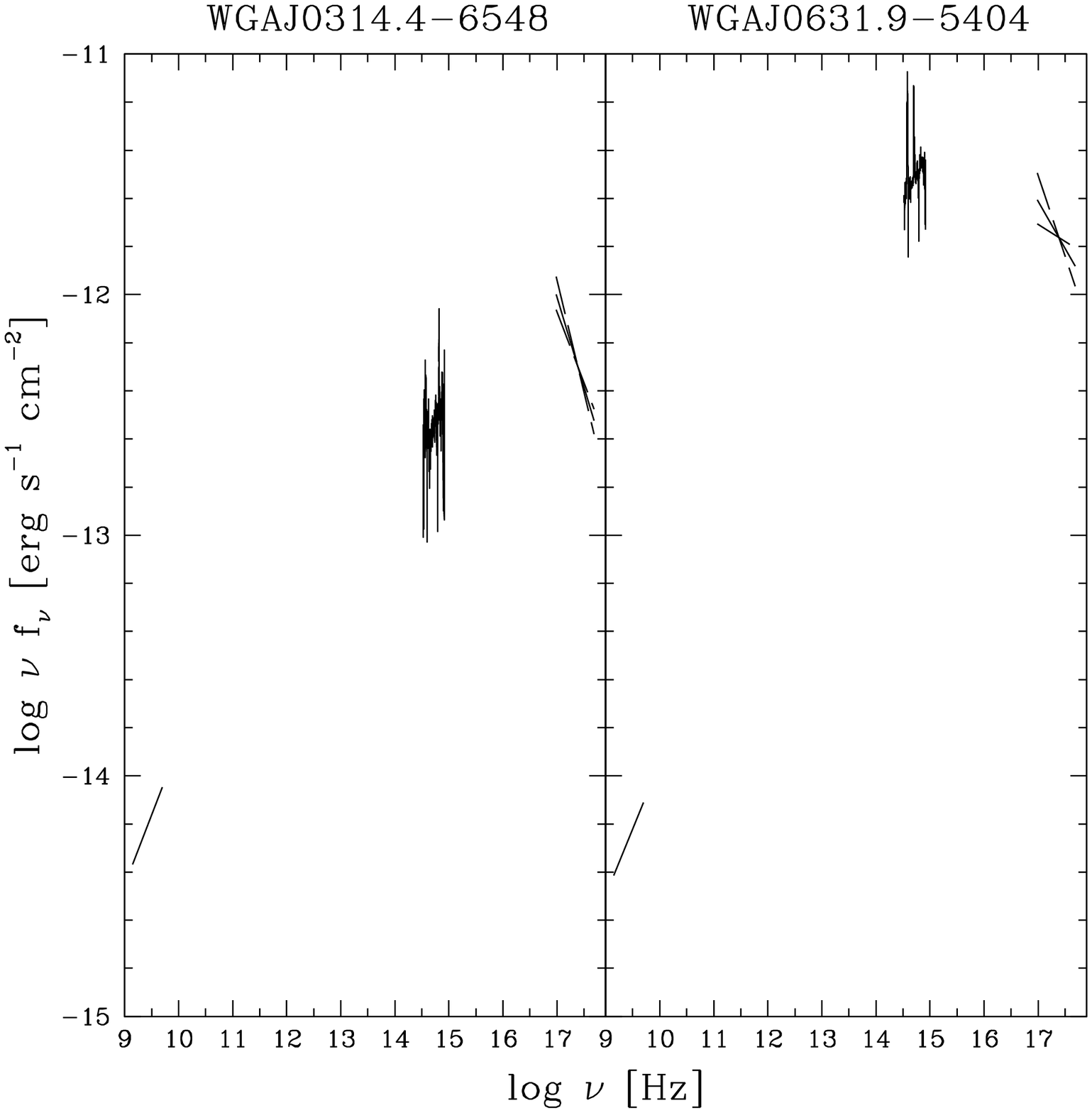,width=3in}
\epsfig{file=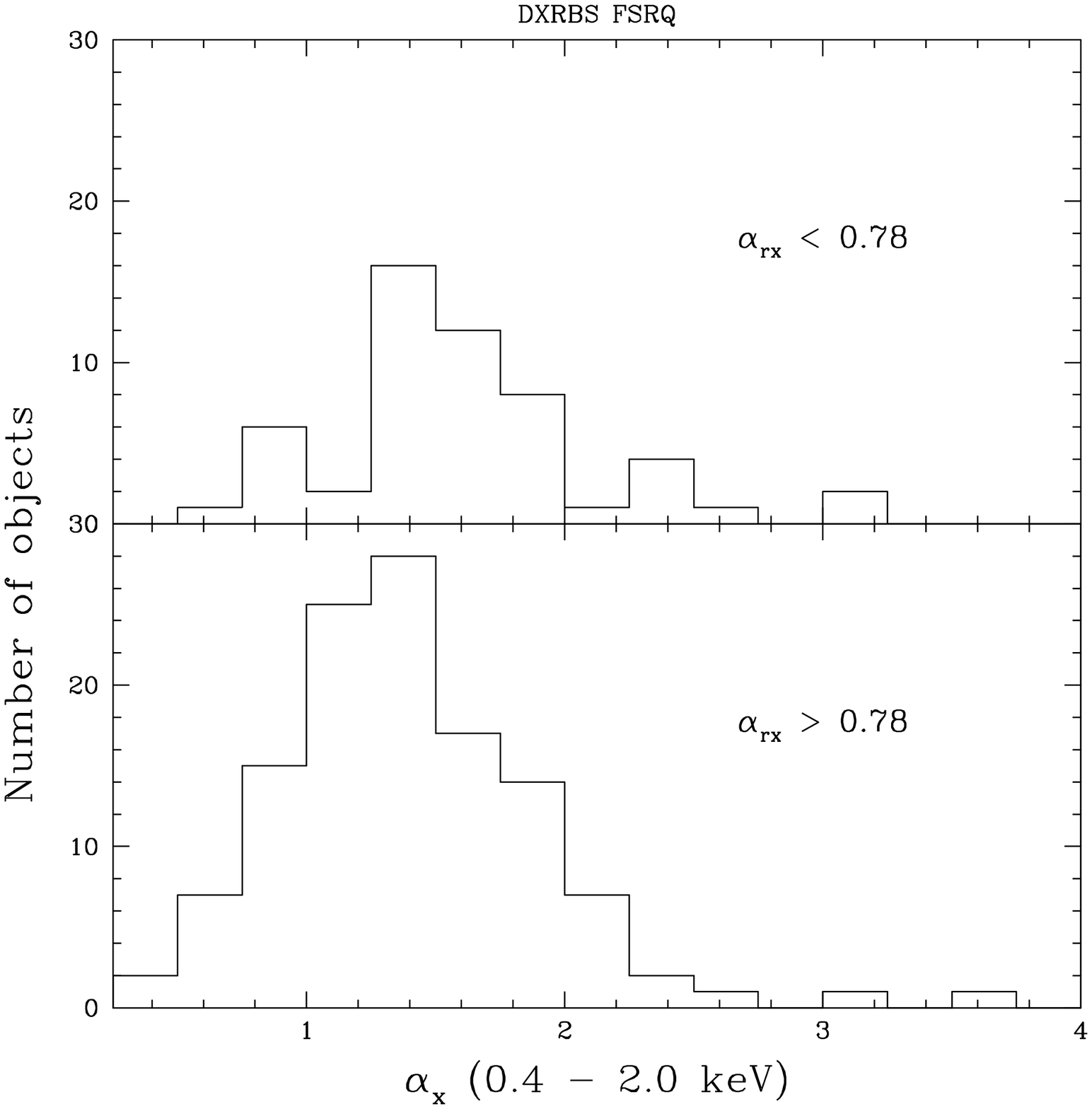, width=3in}}

\caption{At left, the radio to X-ray SED of
two high-energy peaked FSRQ [35].  The data clearly
point to a spectral peak in the UV/X-ray, similar to HBL BL Lacs. 
At right, ROSAT spectral indices of DXRBS FSRQ with radio-to-X-ray spectral
indices above and below 0.78 [35], which is equivalent to the diagonal
line at log $L_x/L_R = 10^{-6}$ on Figure 4.  The spectra of the 
X-ray brighter objects are steeper than those of more radio-loud
``traditional'' FSRQ. }
\end{figure}

This last observation is a fly in the ointment, because the most
natural explanation of the radio to X-ray SEDs derived from survey
data is synchrotron radiation from a single population of electrons,
as with HBL BL Lacs.  If indeed this paradigm holds, one would expect
steeper hard X-ray spectra, dominated by the steep tail of the
particle distribution.  The observation that many of these objects
appear to have flat hard X-ray spectra is difficult to explain in this
context (note that the curvature is in the opposite sense for ``blue
bump'' emission, however).  There are several possibilities.  For
example, if Compton cooling is stronger in these objects (due to
higher electron density, for example) than HBL BL Lacs, Comptonization
could begin to dominate the energy budget at energies of a few keV
instead of tens to hundreds of keV as it does in the HBL BL Lacs.  It
is also possible that there might be some Compton reflected emission
from the accretion disk in the hard X-rays, as there is for Seyfert
galaxies (e.g., \cite{Grandi98}).  However, no spectral curvature is
seen at a few keV \cite{Costa99,Sambruna99}; nor do we see a strong Fe
K$\alpha$ line from the inner reaches of the accretion disk, as is
seen in Seyferts.  There is also a question of selection: Padovani \&
Giommi \cite{GiPa94,PaGi95} showed that (assuming a roughly parabolic
SED), a hypothetical object moves diagonally down on the
($\alpha_{ox},\alpha_{ro}$) plane (roughly along the locus of the 1 Jy
sample) until its peak reaches the optical, after which it begins to
move horizontally to the left (roughly along the locus of the X-ray
selected surveys).  Thus one may need to make two cuts (in
$\alpha_{ro}$ and $\alpha_{rx}$, as below) rather than one in
order to select high-energy peaked objects.

If indeed the radio to X-ray continuum in high-energy peaked FSRQ is
produced by synchrotron radiation, one might expect that Comptonized
emission would peak at around 1 TeV (similar to HBL BL Lacs), so that
these objects should be targets for new TeV observatories.  However,
this assumes that the balance of mechanisms in Compton cooling is not
too different in these objects than it is in HBL BL Lacs.

To produce a list of TeV candidates, I analyzed the DXRBS and RGB
samples, as well as existing surveys, which previously had unknown
numbers of FSRQ.  I found significant numbers of FSRQ in both the EMSS
(16) and Slew (19); About 30\% of these appear to be high-energy
peaking objects (\cite{Perlman99}; 7/16 in the EMSS and 4/19 in the
Slew), consistent with the DXRBS and RGB results \cite{PaPe99}.  I
have selected from these surveys $z<0.1$ objects with
$\alpha_{ro}<0.6$ and $\alpha_{rx}<0.78$ and F(X)$>10^{-12} {\rm ~erg
~cm^{-2} ~s^{-1}}$.  The result is shown in Table 2\footnote{The NVSS-RASS
survey will add a few candidates when their survey is complete; a first 
list of TeV candidates is expected to be released soon in \cite{Giommi00}}.  

\begin{table}[ht!]
\caption{Suggested Candidates for TeV Emission}
\begin{tabular*}{6in}{lcccccl}
Name & RA & Dec & F(0.3--3.5 keV)\tablenote{corrected for galactic absorption}\cite{PaGi97} & $F(R)$ & $z$ & ID \\
     &  \multicolumn{2}{c}{(J2000)}       & $10^{-12}{\rm ~erg ~cm^{-2} ~s^{-1}}$ & mJy & \\
\tableline
1ES0033+595 & 00 35 52.7 & +59 50 04 & 75.1 & 66 & 0.086 & BL Lac \\
RGB0110+418 & 01 10 04.8 & +41 49 50 & 2.5\tablenote{uncorrected 0.1-2.4 keV Flux taken from \cite{L-M99}; no 0.3-3.5 keV flux measurement available} & 40  & 0.096 & BL Lac \\
RGB0152+017 & 01 52 39.7 & +01 47 18 & 5.0$^b$ & 51  & 0.080 & BL Lac \\
RGB0153+712 & 01 53 25.9 & +71 15 07 & 3.1$^b$ & 643 & 0.022 & BL Lac \\
RGB0214+517 & 02 14 17.9 & +51 44 52 & 13.0$^b$ & 291 & 0.049  & BL Lac\\
RGB0314+247 & 03 14 02.7 & +24 44 31 & 2.1$^b$ &  6 & 0.054 & BL Lac \\
1ES0548--322 & 05 50 41.9 & $-$32 16 11 & 44.1  & 170 & 0.069 & BL Lac \\
RGB0656+426 & 06 56 10.7 & +42 37 02 & 3.9$^b$ & 480 & 0.059 & BL Lac \\
Mkn 180 & 11 36 26.4 & +70 09 28 & 7.1 & 94 & 0.046 & BL Lac \\
RGB1532+302 & 15 32 02.2 & +30 16 28 & 5.9$^b$ & 42  & 0.064  & BL Lac \\
RGBJ1610+671 & 16 10 02.6 & +67 10 29 & 4.8$^b$ & 36 & 0.067 & BL Lac \\
1ES1727+502 & 17 28 18.5 & +50 13 11 & 13.7 & 159 & 0.055 & BL Lac \\
1ES1741+196 & 17 43 57.5 & +19 35 10 & 24.6 & 223 & 0.083 & BL Lac \\
1ES1959+650 & 19 59 59.9 & +65 08 55 & 83.4  & 252 & 0.048 & BL Lac \\
PKS2005--489 & 20 09 25.3 & $-$48 49 53 & 16.1 & 1192 & 0.071 & BL Lac\\ 
1ES2321+419 & 23 23 52.0 & +42 11 00 & 2.2 & 19 & 0.059 & BL Lac \\
RGB 2322+346 & 23 22 44.0 & +34 36 14 & 2.2$^b$ & 78 & 0.098 & BL Lac \\
III Zw 2 & 00 10 31.0 & +10 58 28 & 4.8  & 420 & 0.090 & FSRQ \\
B2 0138+398 & 01 41 57.8 & +39 23 30 & 1.1$^b$ & 115 & 0.080 & FSRQ \\
B2 0321+33 & 03 24 41.2 & +34 10 45 & 6.6 & 364 & 0.062 & FSRQ \\
RGB1413+436 & 14 13 43.7 & +43 39 45 & 4.5$^b$ & 39 & 0.090 & FSRQ \\
PG 2209+184 & 22 11 53.7 & + 18 41 51 & 8.4 & 134 & 0.070 & FSRQ \\

\end{tabular*}
\end{table}

Importantly, lists such as Table 2 should {\it not} serve as a be-all
and end-all for future TeV surveys.  This is
particularly true for southern hemisphere observers, since most blazar
surveys unfortunately cover very little of the southern sky!  It is
very important that large angle TeV surveys be carried out in the
near future, and I am encouraged to see that one is now being
planned\cite{Dingus99}.  It has become a truism that every time a
large-area survey is done in a new waveband, some completely
unexpected discovery is made.  It is also equally true that without
large-area surveys, it is not possible to correctly derive physics for
an entire class.  An example of this comes from comparing the ratio of
GeV to radio emission typical for EGRET detected blazars ($\sim 700$)
with that required of all blazars in order to produce the diffuse GeV
background ($\sim 70$;
\cite{KaPe97,SteSa96}). If we based all our modeling of
how the GeV continuum of blazars is produced upon the assumption that
nature only makes blazars with GeV/radio ratios in the hundreds, our
model would not be accurate for the vast majority of sources.  The
same holds in the TeV\cite{Urry99}.

\section{Conclusions}

It is safe to say that there are many key open questions in
blazar research.  Future surveys will be crucial in addressing many of
these.  Here are a few examples:

\begin{itemize}

\item What is the nature of the HBL-LBL relationship (in both BL Lacs
{\it and} FSRQ), and the BL Lac-FSRQ
relationship?  These questions can only be addressed properly when
radio selected samples have large numbers of HBL type objects, and
X-ray selected samples have large numbers of LBL type objects.
Gamma-ray surveys can also help here: each model for the
production of gamma-rays has a different dependence on viewing 
angle and Lorentz $\Gamma$ \cite{Boe99,Rach99,Mast99}.

\item What constrains emission line luminosity and what role does the 
emission line luminosity play in gamma-ray production in various types
of blazars?  One prerequisite for answering this question
now exists:  large surveys which contain blazars of
all X-ray/radio continuum and emission line luminosity classes.  But
the other does not: a gamma-ray database that includes
variability information on statistically significant numbers of
objects at all luminosities.

\item What are the gamma-ray spectral energy distributions of all blazars?
The prerequisite for answering this survey is gamma-ray surveys in the MeV,
GeV and TeV ranges, which are deep enough to contain statistically significant
numbers of blazars of all continuum and emission line luminosity classes.

\item Are blazars (whether aligned or misaligned) really the only 
extragalactic objects which produce detectable quantities of gamma-rays 
with energies greater than $\sim 10$ MeV?   This question is intimately 
related to the issue of what produces the gamma-ray background\cite{KaPe97}.

\end{itemize}

These questions, and the new issues they raise, are intimately
connected with, and equally as important, as those that will be addressed
(and raised) in multiwavelength campaigns and modelling efforts.

\bigskip

This paper summarizes work I have done in collaboration with several
other scientists.   I would particularly like to acknowledge my DXRBS
collaborators, Paolo Padovani, Paolo Giommi, Hermine Landt and Rita 
Sambruna.

\end{document}